\documentclass[showpacs,preprintnumbers,amsmath,amssymb,floatfix]{revtex4}
\usepackage{epsf}
 \newcommand{\insertplot}[5]{\begin{figure}
 \hfill\hbox to 0.05in{\vbox to #5in{\vfill
 \inputplot{#1}{#4}{#5}}\hfill}
 \hfill\vspace{-.1in}
 \caption{#2}\label{#3}
 \end{figure}}
 \newcommand{\inputplot}[3]{
 \special{ps: plotfile #1}
}

\usepackage[german, english]{babel}
\usepackage{ifthen}
\usepackage{epsfig}
\newcounter{fig}   \newcommand{\lbfig}[1]{\refstepcounter{fig}
\label{#1} }

\begin{document}
\title{Electroweak Sphalerons with Spin and Charge}
\author{
{\bf Burkhard Kleihaus, Jutta Kunz and Michael Leissner}
}
\affiliation{
{Institut f\"ur Physik, Universit\"at Oldenburg, 
D-26111 Oldenburg, Germany}
}
\date{\today}

\begin{abstract}
We show that, at finite weak mixing angle
the sphaleron solution of Weinberg-Salam theory
can be endowed with angular momentum
proportional to the electric charge.
Carrying baryon number 1/2 these
sphalerons with spin and charge may contribute to
baryon number violating processes.
\end{abstract}

\maketitle

\section{Introduction}

Like magnetic monopoles, sphalerons arise 
as classical solutions in theories
which undergo spontaneous symmetry breaking.
The Higgs field then defines a map from the two-sphere at infinity
to the vacuum manifold, associated with the symmetry breaking.
But while this mapping is non-trivial 
for monopoles, it is trivial for sphalerons.
Consequently, lacking a conserved topological charge,
sphalerons represent unstable solutions 
associated in a semiclassical approximation 
not with particles but with transition rates. 

In the standard model the electroweak sector 
gives rise to sphalerons 
\cite{Manton:1983nd,Klinkhamer:1984di,Kleihaus:1991ks}.
The static sphaleron solution of Weinberg-Salam theory represents
the top of the energy barrier between topologically inequivalent vacua.
Since the standard model does not absolutely conserve
baryon number \cite{'tHooft:1976up}, at finite temperature
baryon number violating processes can arise because of
thermal fluctuations of the fields
large enough to overcome the energy barrier between 
distinct vacua.
The rate for baryon number violating processes
is then largely determined by a Boltzmann factor,
containing the height of the barrier at a given temperature
\cite{McLerran:1993rv,Rubakov:1996vz,Klinkhamer:2003hz,Dine:2003ax}.
The sphaleron itself carries baryon number $Q_{\rm B}=1/2$ 
\cite{Klinkhamer:1984di}.

At finite weak mixing angle the static electroweak sphaleron
possesses a large magnetic moment but does not carry electric charge.
Adding electric charge to the configuration should lead
to a non-vanishing Poynting vector and thus a finite angular momentum
density of the system, and consequently give rise to
a branch of spinning electrically charged sphalerons.
Carrying non-vanishing baryon number as well,
these configurations would then also entail
baryon number violating processes.

Here we explicitly construct
this branch of spinning electrically charged sphalerons,
and study the dependence on the weak mixing angle.
We show that the angular momentum and the electric charge of the solutions
are proportional \cite{VanderBij:2001nm}.
Electroweak sphalerons thus present the first spinning configuration,
based on non-Abelian gauge fields,
which corresponds to a single localized lump.
Previously, only composite configurations such as monopole-antimonopole
pairs were known to rotate \cite{Paturyan:2004ps,Kleihaus:2005fs},
whereas magnetic monopoles or dyons were shown to
exclude slow rotation \cite{Heusler:1998ec,Volkov:2003ew}.

In section 2 we present the action, the Ansatz and the
boundary conditions. In section 3 we consider the
relevant physical properties and, in particular, derive the linear relation 
between angular momentum and electric charge. 
We present and discuss the numerical results in section 4.

\section{Action and Ansatz}


We consider the bosonic sector of Weinberg-Salam theory
\begin{equation}
{\cal L} = -\frac{1}{2} {\rm Tr} (F_{\mu\nu} F^{\mu\nu})
-  \frac{1}{4}f_{\mu \nu} f^{\mu \nu}                                           
- (D_\mu \Phi)^{\dagger} (D^\mu \Phi) 
- \lambda (\Phi^{\dagger} \Phi - \frac{v^2}{2} )^2 
\  
\label{lag1}
\end{equation}
with su(2) field strength tensor
\begin{equation}
F_{\mu\nu}=\partial_\mu V_\nu-\partial_\nu V_\mu
            + i g [V_\mu , V_\nu ]
 , \end{equation}
su(2) gauge potential $V_\mu = V_\mu^a \tau_a/2$,
u(1) field strength tensor
\begin{equation}
f_{\mu\nu}=\partial_\mu A_\nu-\partial_\nu A_\mu 
 , \end{equation}
and covariant derivative of the Higgs field
\begin{equation}
D_{\mu} \Phi = \Bigl(\partial_{\mu}
             +i g  V_{\mu} 
             +i \frac{g'}{2} A_{\mu} \Bigr)\Phi
 , \end{equation}
where $g$ and $g'$ denote the $SU(2)$ and $U(1)$ gauge coupling constants,
respectively,
$\lambda$ denotes the strength of the Higgs self-interaction and
$v$ the vacuum expectation value of the Higgs field.

The gauge symmetry is spontaneously broken 
due to the non-vanishing vacuum expectation
value of the Higgs field
\begin{equation}
    \langle \Phi \rangle = \frac{v}{\sqrt2}
    \left( \begin{array}{c} 0\\1  \end{array} \right)   
 , \label{Higgs} \end{equation}
leading to the boson masses
\begin{equation}
    M_W = \frac{1}{2} g v   , \ \ \ \ 
    M_Z = \frac{1}{2} \sqrt{(g^2+g'^2)} v  \, , \ \ \ \ 
    M_H = v \sqrt{2 \lambda} \,  . 
\end{equation}
$ \tan \theta_{\rm w} = g'/g $ determines
the weak mixing angle $\theta_{\rm w}$,
defining the electric charge $e = g \sin \theta_{\rm w}$.  


To obtain stationary rotating solutions of the bosonic sector
of Weinberg-Salam theory,
we employ the time-independent axially symmetric Ansatz
\begin{equation}
V_\mu\, dx^\mu
  = \left( B_1\, \frac{\tau_r}{2g} + B_2\, \frac{\tau_\theta}{2g} \right) \, dt 
            -\sin\theta\left(H_3 \frac{\tau_r}{2g}
            + H_4 \frac{\tau_\theta}{2g}\right)\, d\varphi
+\left(\frac{H_1}{r}\, dr +(1-H_2)\, d\theta \right)\frac{\tau_\varphi}{2g}
  , \label{a1} \end{equation}
\begin{equation}
A_\mu\, dx^\mu = \left( a_1\, dt + a_2\, \sin^2 \theta \, d\varphi \right)/g'
  , \end{equation}
and
\begin{equation}
\Phi = i\, \phi \, ( \cos\psi\, \tau_r + \sin\psi\, \tau_\theta )
    \frac{v}{\sqrt2} \left( \begin{array}{c} 0\\1  \end{array} \right) 
  . \end{equation}
where $\tau_r$ denotes the cartesian product of the Pauli matrices
and the spherical unit vector $e_r$, etc.
The ten functions $B_1$, $B_2$, $H_1,\dots,H_4$, $a_1$, $a_2$,
$\phi$, and $\psi$ depend on $r$ and $\theta$, only. 
%
With this Ansatz the full set of field equations reduces to a system 
of ten coupled partial differential equations in the independent variables 
$r$ and $\theta$. A residual $U(1)$ gauge degree of freedom is 
fixed by the condition $r\partial_r H_1 - \partial_\theta H_2=0$ 
\cite{Kleihaus:1991ks}.

Requiring regularity and finite energy, we impose the boundary conditions 
\begin{eqnarray}
r=0: &  & 
B_1 \sin \theta + B_2 \cos \theta =0  , \
\partial_r \left( B_1 \cos \theta - B_2 \sin \theta\right) =0  , \
H_1=H_3=H_4=0 , \ H_2=1 , \ 
\partial_r a_1 =0  , \ a_2=0  , \
\nonumber \\
&  & 
\phi=0  , \ \partial_r \psi =0
\nonumber \\
r\rightarrow \infty: &  &  
B_1 = \gamma \cos \theta  , \ B_2 = \gamma \sin \theta  , \
H_1=H_3=0 , \ H_2=-1  ,  \ H_4=2 , \ 
a_1=\gamma  , \ a_2 = 0  , \
\phi = 1  , \ \psi = 0
\nonumber \\
\theta = 0: &  & 
\partial_\theta B_1 =0  , \ B_2 =0  , \
H_1=H_3=0 , \ \partial_\theta H_2=\partial_\theta H_4=0 , \ 
\partial_\theta a_1 = \partial_\theta a_2 = 0  , \
\partial_\theta \phi = 0  , \ \psi=0  ,
\end{eqnarray}
where the latter hold also at $\theta=\pi/2$,
except for $B_1=0$ and $\partial_\theta B_2 =0$.
Here $\gamma$ is a constant.

\section{Sphaleron properties}

We now address the global charges of the sphaleron solution,
its mass, angular momentum, electric charge, and baryon number.
The mass $M$ and angular momentum $J$ are defined in terms of
volume integrals of the respective components
of the energy-momentum tensor. The mass is obtained from
\begin{equation}
M = \int T_t^t d^3 r , 
\label{Mint}
\end{equation}
while the angular momentum
\begin{equation}
J = \int T_\varphi^t d^3 r 
= \int \left[ 2 {\rm Tr} \left( F^{t\mu} F_{\varphi\mu} \right)
 + f^{t\mu} f_{\varphi\mu} + 2 \left( D^t \Phi \right)^\dagger
 \left( D_\varphi \Phi \right)   \right] d^3 r
\label{Jint}
\end{equation}
can be reexpressed with help of the
equations of motion and the symmetry properties of the Ansatz
\cite{VanderBij:2001nm,Kleihaus:2002tc,Volkov:2003ew}
as a surface integral at spatial infinity 
\begin{equation}
J =  \int_{S_2} \left\{ 
 2 {\rm Tr} \left( \left(V_\varphi - \frac{\tau_z}{2g}\right) F^{r t} \right) 
 + 
 \left(A_\varphi - \frac{1}{g'}\right) f^{r t}
 \right\} r^2 \sin \theta d \theta d \varphi .
\label{Joint}
\end{equation}
The power law fall-off of the $U(1)$ field
of a charged solution allows for a finite flux
integral at infinity and thus a finite angular momentum.
Insertion of the asymptotic expansion for the $U(1)$ field
\begin{eqnarray}
a_1= \gamma + \frac{Q}{r} + O \left(\frac{1}{r^2}\right) , \nonumber\\
a_2= -\frac{\mu}{r} + O \left(\frac{1}{r^2}\right) ,
\label{asymp}
\end{eqnarray}
and of the analogous expansion for the $SU(2)$ fields
then yields for the angular momentum
\begin{equation}
\frac{J}{4 \pi} = 
   \frac{Q}{g^2} + \frac{Q}{{g'}^2}
= \frac{Q}{g^2 \sin^2 \theta_{\rm w}} = \frac{Q}{e^2} .
\label{Joint2}
\end{equation}

The field strength tensor ${\cal F}_{\mu\nu}$ of the 
electromagnetic field ${\cal A}_{\mu}$,
\begin{equation}
{\cal A}_{\mu}=\sin \theta_{\rm w} V^3_\mu + \cos \theta_{\rm w} A_\mu ,
\label{emfield}
\end{equation}
as given in a gauge where the Higgs field asymptotically tends to
Eq.~(\ref{Higgs}),
then defines the electric charge $\cal Q$
\begin{equation}
 {\cal Q} =  \frac{1}{4\pi} \int_{S_2}
  {^*}{\cal F}_{\theta\varphi} d\theta d\varphi
 = \frac{\sin \theta_{\rm w} Q}{g} + \frac{\cos \theta_{\rm w} Q}{g'}
 = \frac{Q}{e}
  , \label{Qel} \end{equation}
where the integral is evaluated at spatial infinity.
%
Comparison of Eqs.~(\ref{Joint2}) and (\ref{Qel}) then
yields a linear relation between the angular momentum $J$
and the electric charge $\cal Q$
\begin{equation}
 J = \frac{4 \pi \cal Q}{e}
\label{JQrel}
 . \end{equation}
The magnetic moment $\mu$ is obtained from 
the asymptotic expansion Eq.~(\ref{asymp}), analogously to the 
electric charge.

Addressing finally the baryon number $Q_{\rm B}$,
its rate of change is given by
\begin{equation}
 \frac{d Q_{\rm B}}{dt} = \int d^3 r \partial_t j^0_{\rm B}
= \int d^3 r \left[ \vec \nabla \cdot \vec j_{\rm B}
 + \frac{1}{32 \pi^2} \epsilon^{\mu\nu\rho\sigma} \, \left\{
 g^2 {\rm Tr} \left(F_{\mu\nu} F_{\rho\sigma} \right) 
 + \frac{1}{2} {g'}^2 f_{\mu\nu} f_{\rho\sigma} \right\} \right]  . 
\end{equation}
Starting at time $t=-\infty$ at the vacuum with $Q_{\rm B}=0$,
one obtains the baryon number of a sphaleron solution at
time $t=t_0$ \cite{Klinkhamer:1984di},
\begin{equation}
 Q_{\rm B} = 
\int_{-\infty}^{t_0} dt \int_S \vec K \cdot d \vec S
+  \int_{t=t_0} d^3r K^0 
  , \end{equation}
where the $\vec \nabla \cdot \vec j_{\rm B}$ term is neglected,
and the anomaly term is reexpressed in terms of the
Chern-Simons current
\begin{equation}
 K^\mu=\frac{1}{16\pi^2}\varepsilon^{\mu\nu\rho\sigma} 
 \left\{ g^2 {\rm Tr}\left( F_{\nu\rho}V_\sigma
 - \frac{2}{3} i g V_\nu V_\rho V_\sigma \right)
 + \frac{1}{2} {g'}^2 f_{\nu\rho}A_\sigma \right\}
  . \end{equation}
In a gauge, where
\begin{equation}
V_\mu \to \frac{i}{g} \partial_\mu \hat{U} \hat{U}^\dagger   , \ \ \ 
\hat{U}(\infty) = 1   , 
\end{equation}
$\vec K$ vanishes at infinity. 
Subject to the above ansatz and boundary conditions
the baryon charge
of the sphaleron solution \cite{Kleihaus:1994tr} is then
\begin{equation}
 Q_{\rm B} = \int_{t=t_0} d^3r K^0  = \frac{1}{2} .
\label{Q}
\end{equation}

\section{Results and discussion}


We solve the set of ten coupled non-linear
elliptic partial differential equations numerically \cite{FIDISOL},
subject to the above boundary conditions
in compactified dimensionless coordinates,
$x = \tilde r/(1+ \tilde r)$, with $\tilde r = gvr$.

Employing the physical value for the mixing angle $\theta_{\rm w}$,
and increasing the asymptotic value of the $U(1)$ field 
$\tilde \gamma = \gamma/gv$,
a branch of rotating charged sphalerons emerges
smoothly from the static sphaleron.
The branch ends when a limiting value $\tilde \gamma_{\rm max}$
is reached \cite{footnote}. Here some of the gauge field functions
no longer decay exponentially,
precluding localized solutions for larger values of $\tilde \gamma$.
At $\tilde \gamma_{\rm max}$ the solution has maximal spin, charge and mass.
The dependence of the angular momentum $J$ on $\gamma$
is illustrated in Fig.~\ref{f-1}.
The figure also demonstrates the linear relation 
(\ref{JQrel}) between the charge $\cal Q$ and
the angular momentum $J$.
In Fig.~\ref{f-2} we exhibit the dependence of the mass $M$ 
and the magnetic moment $\mu$ on the angular momentum $J$.
\begin{figure}[h!]
\lbfig{f-1}
\begin{center}
\includegraphics[height=.25\textheight, angle =0]{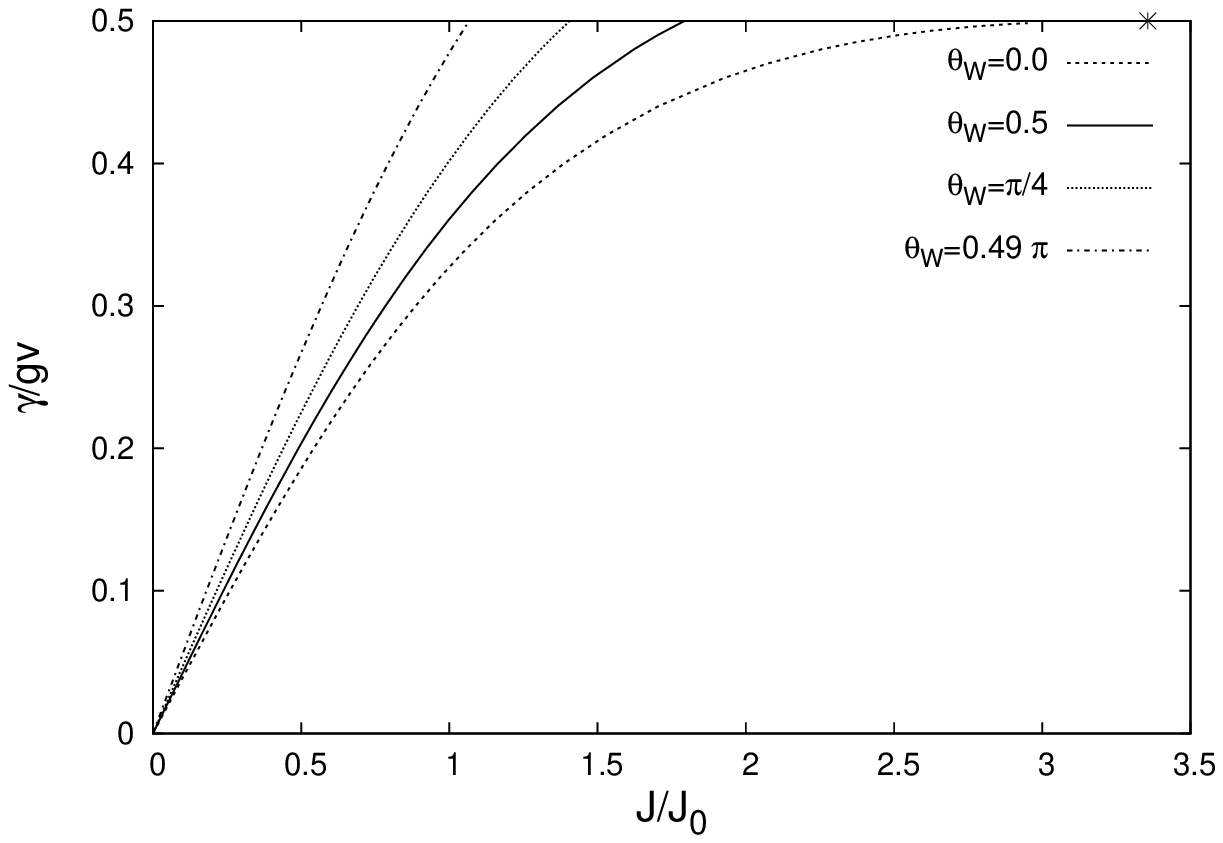}
\includegraphics[height=.25\textheight, angle =0]{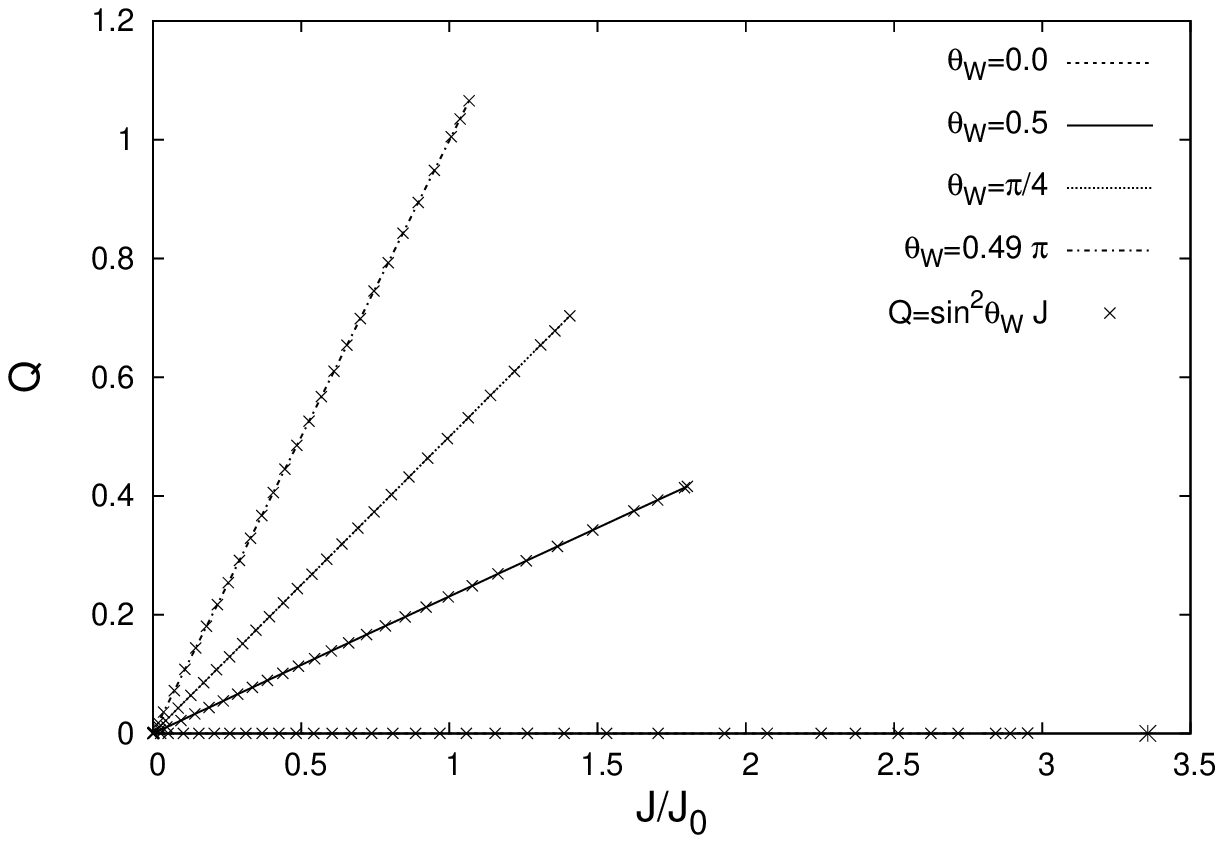}
\end{center}
\caption{
(a) The asymptotic value of the $U(1)$ field $\tilde \gamma = \gamma/gv$ 
and (b) the $U(1)$ charge $Q$ versus the
angular momentum $J$ ($J_0=4 \pi/g^2$)
for several values of the mixing angle $\theta_{\rm w}$.
The asterisk marks the extrapolated maximal value for $J$
for $\theta_{\rm w}=0$.
}
\end{figure}
%
\begin{figure}[h!]
\lbfig{f-2}
\begin{center}
\includegraphics[height=.25\textheight, angle =0]{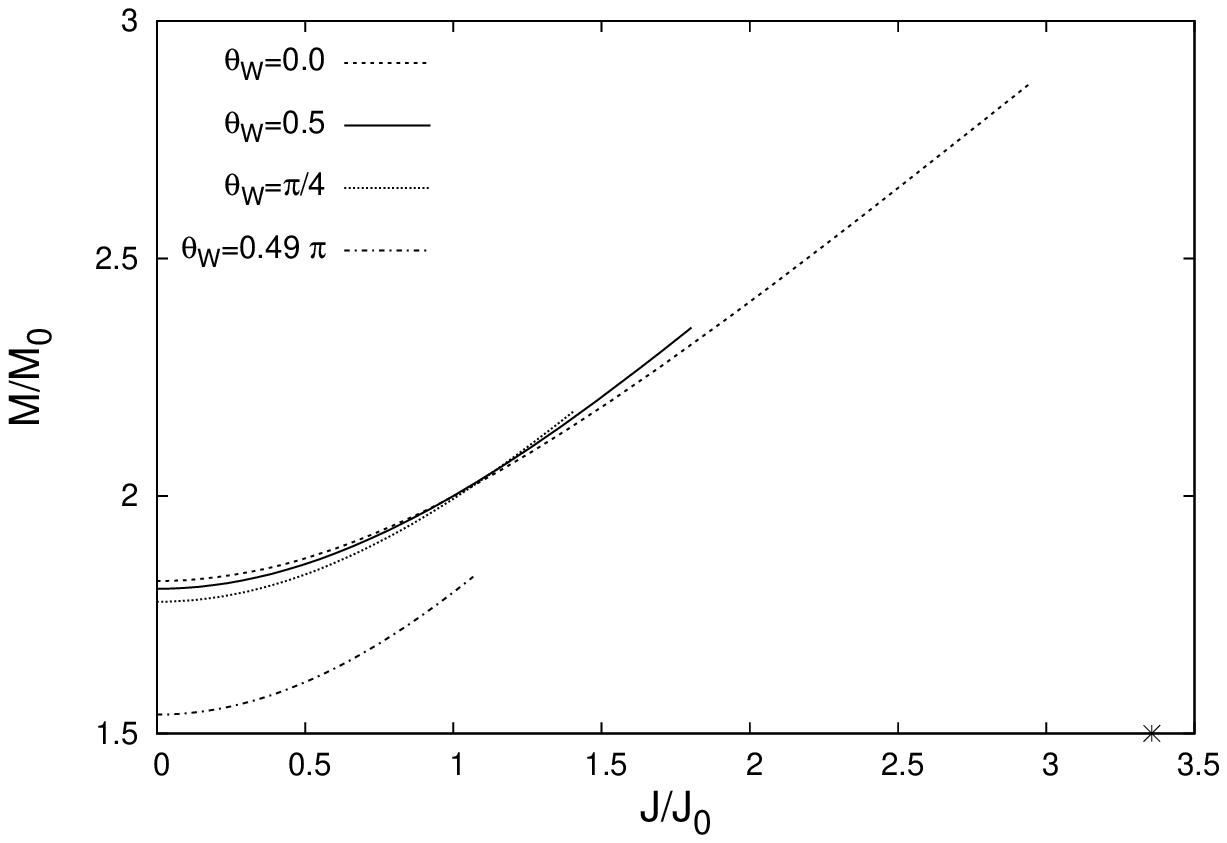}
\includegraphics[height=.25\textheight, angle =0]{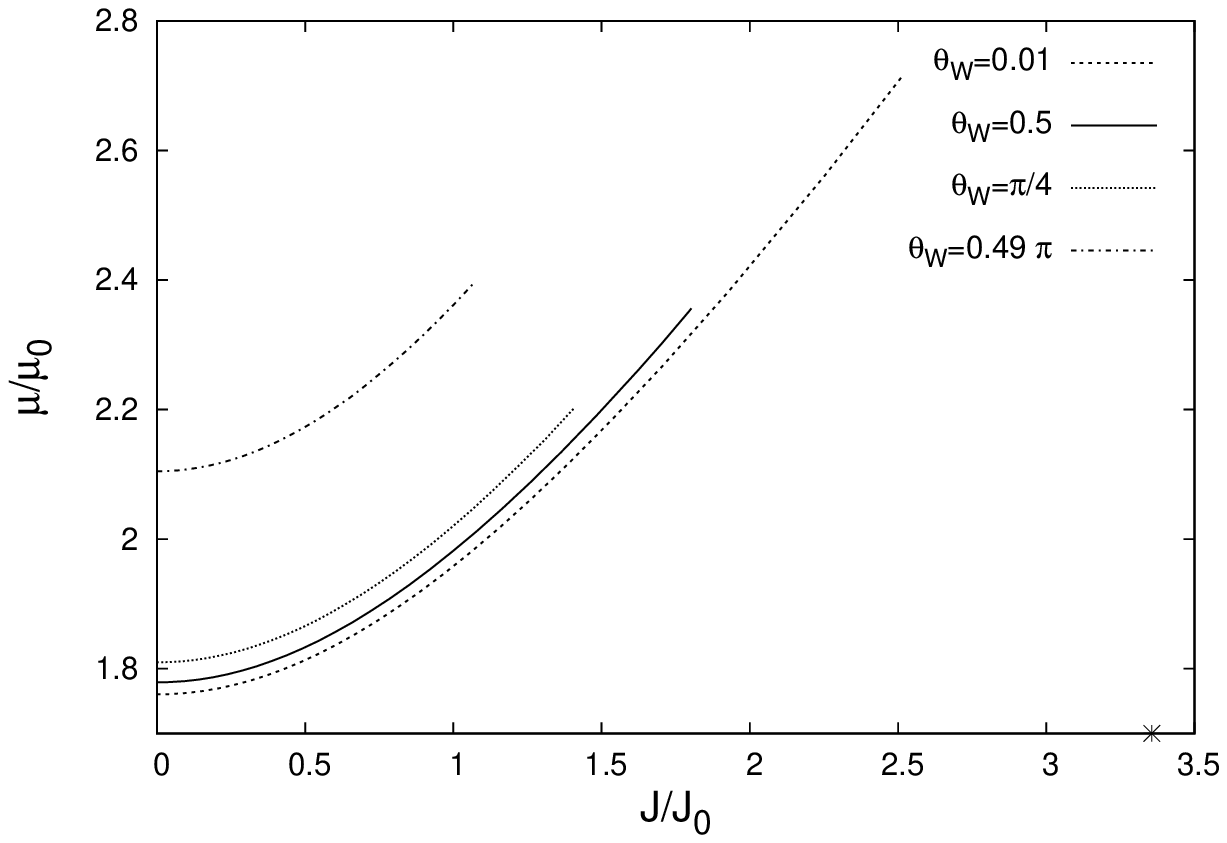}
\end{center}
\caption{
Same as Fig.~\ref{f-1} for
(a) the mass $M$ ($M_0=v/g$) and (b) the magnetic moment $\mu$ 
($\mu_0=2 \sin^2 \theta_{\rm w}/gv$).
}
\end{figure}

To gain further understanding of these rotating solutions,
we now address the mixing angle dependence,
varying $\theta_{\rm w}$ in the range $0 \le \theta_{\rm w} \le \pi/2$.
We recall that, as $\theta_{\rm w}$ is increased, 
the mass of the static sphaleron decreases, and its energy density, 
being spherical at $\theta_{\rm w}=0$, becomes increasingly prolate
\cite{Kunz:1992uh}.

Choosing a fixed value of $\theta_{\rm w}$ beyond the physical value,
the respective branch of rotating charged solutions
exhibits a slower increase of angular momentum $J$ with 
$U(1)$ parameter $\gamma$, and thus a smaller maximal value of $J$.
In the limit $\theta_{\rm w} \rightarrow  \pi/2$,
the relation between $J$ and $\gamma$ becomes almost linear,
as seen in Fig.~\ref{f-1}, for $\theta_{\rm w}=0.49 \pi$.
Note, that the limit cannot be obtained numerically.

Let us now consider smaller values of $\theta_{\rm w}$, and in particular
the limit $\theta_{\rm w} \rightarrow 0$.
In this limit the sphaleron was expected not to rotate \cite{Volkov:2003ew}.
The numerical data, as shown in Fig.~\ref{f-1}, however indicate
the presence of a rotating branch of solutions in this limit.
In fact, for a given value of $\gamma$, 
the angular momentum increases with decreasing $\theta_{\rm w}$,
assuming its largest value in the limit.
The charge on the other hand, decreases to zero in this limit,
thus provoking the question as to 
what then allows for the rotation?

Analysis then shows that the $U(1)$ field becomes trivial
in the limit $\theta_{\rm w} \rightarrow 0$,
except that $a_1$ assumes a finite constant value, $a_1=\gamma$.
Thus $\gamma$ enters the covariant derivative of the Higgs field,
and provides non-trivial boundary conditions for the time-component
of the $SU(2)$ gauge field, thus giving rise to a non-vanishing
$SU(2)$ Poynting vector and, consequently, angular momentum.

These limiting solutions can also be considered from an alternative
point of view. Giving the Higgs field a time-dependent phase,
as discussed in \cite{Radu:2008pp}, 
$\gamma$ enters as a frequency parameter,
analogous as in non-topological solitons.
This permits an otherwise identical set of solutions with the same
$\gamma$-dependence of the angular momentum and the mass.

Let us finally consider the effect of spin and charge
on the energy density and thus the shape of the configuration.
The effect of charge is to spread the energy density further out, 
while reducing its central magnitude, 
as also seen in dyons, for instance.
This effect is quite pronounced for large charge.
On the other hand,
the expected effect of angular momentum, i.e., 
a relative centrifugal flattening of the shape of the energy density,
is barely noticable even at maximal spin.
In particular, for larger $\theta_{\rm w}$, 
the prolate deformation of the solutions is retained,
and only marginally reduced in the presence of rotation.


Concluding, we have shown that the static electroweak sphaleron
gives rise to a branch of rotating electrically charged solutions,
whose angular momentum and charge are proportional.
Carrying baryon number $Q_{\rm B}=1/2$,
they can be associated with baryon number violating processes.
Their presence may thus affect the calculations 
of the generation of the baryon number asymmetry of the universe 
within the standard model 
\cite{McLerran:1993rv,Rubakov:1996vz,Dine:2003ax}.
The inclusion of rotation and charge in more general
solutions of Weinberg-Salam theory, such as multisphalerons
or sphaleron-antisphaleron systems \cite{Kleihaus:2008gn},
is currently under study. 
Further insight is expected from the study
of the fermion modes in the background of these solutions
\cite{Kunz:1993ir}.

{\bf Acknowledgement}:
B.K.~gratefully acknowledges support by the DFG.


\begin{thebibliography}{000}
\bibitem{footnote}
Assuming the exponent to arise from a combination of the 
non-Abelian interaction and mass term,
a decay constant proportional to $\sqrt{1-4\tilde \gamma^2}$
should appear, yielding $\tilde \gamma_{\rm max}=1/2$.
\end{thebibliography}
\end{document}